\documentclass[journal,compsoc,a4paper]{IEEEtran}
\usepackage{amsmath,amsfonts}
\usepackage{multirow}
\usepackage{array}
\usepackage[caption=false,font=normalsize,labelfont=sf,textfont=sf]{subfig}
\usepackage{textcomp}
\usepackage{stfloats}
\usepackage{url}
\usepackage{verbatim}
\usepackage{graphicx}
\usepackage{cite}
\usepackage{makecell}  

\usepackage{listings}
\usepackage{xparse}
\usepackage{xcolor}
\lstdefinestyle{py}{
  language=Python,
  basicstyle=\small\ttfamily,
  commentstyle=\color{green!40!black},
  keywordstyle=\color{blue},
  numberstyle=\tiny\color{gray},
  numbers=left,
  numbersep=5pt,
  backgroundcolor=\color{white},
  showspaces=false,
  showstringspaces=false,
  frame=single,
  rulecolor=\color{black},
  tabsize=4,
  captionpos=b,
  breaklines=true,
  breakatwhitespace=true,
  title=\lstname,
  caption=\lstname
}

\usepackage{algorithm,algpseudocode}

\usepackage[T1]{fontenc} 

\usepackage{hyperref}

\begin{document}

\title{Network Diffusion --– Framework to Simulate Spreading Processes in Complex Networks}

\author{Micha{\l} Czuba\textsuperscript{*1}, Mateusz Nurek\textsuperscript{1}, Damian Serwata\textsuperscript{1}, Yu-Xuan Qiu\textsuperscript{2}, Mingshan Jia\textsuperscript{2}, Katarzyna Musial\textsuperscript{2}, Rados{\l}aw Michalski\textsuperscript{1}, Piotr Br{\'o}dka\textsuperscript{1}.
\thanks{* corresponding author}
\thanks{[1]
are with Department of Artificial Intelligence, Wroc{\l}aw University of Science and Technology, 27 wybrze{\.z}e Wyspia{\'n}skiego st, 50-370 Wroc{\l}aw, Poland, e-mails: {\tt \{michal.czuba, mateusz.nurek, damian.serwata, radoslaw.michalski, piotr.brodka\}@pwr.edu.pl}}
\thanks{[2]
are with Data Science Institute, University of Technology Sydney, PO Box 123,
Ultimo NSW 2007, Australia, e-mails: {\tt \{yuxuan.qiu, mingshan.jia, katarzyna.musial-gabrys\}@uts.edu.au}}

}

\markboth{BIG DATA MINING AND ANALYTICS, ISSN 2096-0654, DOI: 10.26599/BDMA.2024.9020010}{}
\maketitle

\begin{abstract}
With the advancement of computational network science, its research scope has significantly expanded beyond static graphs to encompass more complex structures. The introduction of streaming, temporal, multilayer, and hypernetwork approaches has brought new possibilities and imposed additional requirements. For instance, by utilising these advancements, one can model structures such as social networks in a much more refined manner, which is particularly relevant in simulations of the spreading processes. Unfortunately, the pace of advancement is often too rapid for existing computational packages to keep up with the functionality updates. This results in a significant proliferation of tools used by researchers and, consequently, a lack of a universally accepted technological stack that would standardise experimental methods (as seen, e.g. in machine learning). This article addresses that issue by presenting an extended version of the Network Diffusion library. First, a survey of the existing approaches and toolkits for simulating spreading phenomena is shown and then, an overview of the framework functionalities. Finally, we report four case studies conducted with the package to demonstrate its usefulness: the impact of sanitary measures on the spread of COVID-19, the comparison of information diffusion on two temporal network models, and the effectiveness of seed selection methods in the task of influence maximisation in multilayer networks. We conclude the paper with a critical assessment of the library and the outline of still awaiting challenges to standardise research environments in computational network science.

\end{abstract}


\section{Introduction}

With the rapid growth of data generated by humankind in recent decades, the issue of its efficient processing became one of the main challenges that need to be addressed to boost experimental sciences. That especially concerns irregular structures like graphs that are much harder to enclose in robust frameworks such as images, audio, or text. Computational network science~\cite{menczer2020first}, the primary discipline focusing on applications of graphs, developed many tools that facilitate research-oriented computations. They are used in tasks like epidemiology~\cite{pastor-satorras2001epidemic}, social network analysis~\cite{nurek2023hawkes}, recommendation systems~\cite{forouzandeh2023recommendataionreschet, rostami2022foodrecommendataionimage} or urban planning~\cite{zhong2014detecting} to name just a few. However, the variety and versatility of those solutions significantly stand out from frameworks designed to deal with other data modalities.

As stands in~\cite{Shashank2022InfluenceMaximization}, the number of spreading models and problems they tackle is overwhelming and exceeds the abovementioned examples. One can add to that various network models and forms of simultaneous coexistence of many processes~\cite{brodka2020interacting}. When these phenomena are up to be analysed, researchers must first design and perform adequate simulations. Unfortunately, as demonstrated in our previous work~\cite{czuba2022networkdiffusion}, there is a big proliferation of environments that can be used for that. Hence, outcomes of such research are often presented only as articles without proper code attached, which could allow reproducing results and easily include proposed methods (e.g. a new centrality measure, spreading model, etc.) in following research conducted by another organisation. Although challenges related to the standardisation of experimental environments and results reproducibility have been addressed a long time ago in areas of computer science like computer vision, signal processing, or machine learning, they are still valid in network science.

This paper presents an enhanced version of Network Diffusion, a framework originally designed to simulate the coexisting spreading processes in multilayer networks~\cite{czuba2022networkdiffusion}. Since publication, it has been transformed into a comprehensive library that allows the processing of both temporal and multilayer networks with various spreading processes on top of them. Though the tool has been developed during various research activities of the authors, we collected developed methods and models (most of them are referenced in Tab.~\ref{tab:abbrs}) into one consistent entity. To the best of our knowledge, there is no such library providing similar functionalities. Therefore, this framework can be considered as complementary to other available solutions. We can summarise a contribution brought with this paper as an introduction of a research environment for simulating spreading processes in complex networks to promote open science and research reproducibility. Moreover, with the framework, we tackle and solve (in a minimal way) still valid problems from a domain of influence maximisation, epidemiology, and network modelling.

The outline of the article is the following. First, we present a review of existing solutions tackling given problems (Sec.~\ref{sec:tools}), then the general outline of library functionalities (Sec.~\ref{sec:features}). In Sec.~\ref{sec:examples}, we show four use cases of experiments conducted with the package. After that, in Sec.~\ref{sec:efficiency}, we describe a study of the computational efficiency of Network Diffusion. The paper is summarised in Sec.~\ref{sec:conclusions}. For the reader's convenience, we attach a table with abbreviations used in the article (Tab.~\ref{tab:abbrs}). Names of datasets used in the experimental part can be found in Tab.~\ref{tab:networks_eda}.

\begin{table}[ht]
    \centering
    \caption{Abbreviations used in the paper.}
    \begin{tabular}{l|p{2.5cm}|p{3.5cm}}
 Abbr. & Abbr. expansion & Note \\ \hline
 SIS & Suspected-Infected-Suspected & A basic epidemiological spreading model~\cite{barabasi2016network} \\ \hline
 SIR & Suspected-Infected-Removed & A basic epidemiological spreading model~\cite{barabasi2016network}  \\ \hline
 LTM & Linear Threshold Model & A spreading model in the sense as in~\cite{granovetter1978threshold, kempe2003maximizing}, with extension to multilayer networks as in~\cite{zhong2022mltm} \\ \hline
 ICM & Independent Cascade Model & A spreading model in the sense as in \cite{kempe2003maximizing}, with extension to multilayer networks similarly to~\cite{zhong2022mltm} \\ \hline
 SIR-UA & Suspected-Infected\hspace{1em} $\sim$  Unaware-Aware & Two coexisting spreading models in the sense as in \cite{czuba2022networkdiffusion, brodka2020interacting} \\ \hline
 NEM & Network Epistemology Model & A spreading model in the sense as in~\cite{zollman2007communication}, with extension to temporal networks as in~\cite{michalski2022epistemology} \\ \hline
 MDS & Minimal Dominating Set & A set of network's nodes allowing to control its entire structure according to~\cite{nacher2012dominating, sadaf2023maximising} \\ \hline
 CogSNet & Cognition-Driven Social Network & A temporal network model according to~\cite{michalski2021social} \\
    \end{tabular}
    \label{tab:abbrs}
\end{table}

\section{Existing tools}
\label{sec:tools}

In this section, we would like to briefly introduce existing tools and software packages with functionalities similar to the Network Diffusion package. The detailed list and comparison are already available~\cite{czuba2022networkdiffusion}. Here, we would like to focus on tools that can be used for research on various spreading processes in multilayer and temporal networks.

\subsection{GLEaMviz}
The first application with functionalities corresponding to the designed software is a GLEaMviz~\cite{gleam}. It works with real data, population density, and migration around the world, combined with stochastic models of disease propagation. As a result, it provides a sophisticated simulation environment. Due to the large scale of the experiments (the whole world), a single node is a population of a given size (defined by the user). A very interesting feature is the manual definition of the epidemiological model. GLEaMviz makes this possible by manipulating the compartments. Allowable transitions between them are also fully definable. Users are also able to choose the geographical origin of the disease, specify the initial proportions of individuals in each compartment, determine its duration, and more. There is also an option to generate various visualisations at the end of the experiment. Despite the interesting functionalities mentioned above, GLEaMviz has a few disadvantages: it is limited to disease spreading, only allows the propagation of one process at a time and does not support multilayer and temporal networks.

\subsection{NDlib}\label{NDLIB}
Network Diffusion LIBrary~\cite{NDlib} is a Python package based on the NetworkX library. It allows performing simulations with many predefined epidemiological (such as SIS, SIR, etc.), influence spread (LTM, ICM, Profile, etc.) or opinion formation (Voter, Sznajd, etc.) models, and even dynamics (models with the capacity to change the topology of network). Moreover, the user can create its own customised models. Results visualisation is also possible via Matplotlib or Bokeh with the flexibility to append a custom graphical engine. NDlib also has some interesting run-time features. First, it includes an option to perform a "multi-execution" of the simulation by parallel computing. As this kind of experiment is generally stochastic, this feature gives a chance to see the general behaviour of the observed phenomena. It also enables running the simulation on a server (as well as locally). For users unfamiliar with Python, NDlib's authors created NDQL, a query language (based on SQL syntax) that supports elementary commands of the library. For those who cannot programme, they also provided a "visualisation framework" to play with some of the models implemented with a graphical user interface-based tool. NDlib is a very useful library with many features. However, it does not directly support experiments where multiple spreading processes interact with each other, as well as experiments on temporal networks.

\subsection{SimInf}
The next tool is SimInf - a Framework for Data-Driven Stochastic Disease Spread Simulations~\cite{siminf}, a package for R language. This tool allows a user to define custom spreading models that are executed following a principle of continuous-time Markov chains and the Gillespie stochastic simulation algorithm. Interestingly, no network is needed to run the simulation, as the algorithms work on the assumption of homogeneous mixing and additional data such as births, deaths, and population movements at predefined time points. After a successful simulation, it is possible to display a summary graph. Nevertheless, we must note that no support for operations on real, multilayer and temporal networks, as well as the lack of an option to define interacting processes, are its distinct shortcomings.

\subsection{Sispread}
Sispread~\cite{sispread} is a simple application implemented in C. It is a console tool without a graphical interface. It focuses on epidemic models (SI, SIS, SIR) with the possibility of deep analysis of the performed experiments. This tool supports very basic IO operations - the user can upload a custom network that meets certain requirements (without the option to manipulate it) or generate it using one of three available models: Barabási-Albert, Erdös-Renyi, or Kleinberg. As a result of the experiment, numerical data is returned for further analysis. Similarly to previous packages, it does not support multiple spreading processes as well as multilayer or temporal networks. 

\subsection{STEM}
Another tool is Spatio-Temporal Epidemiological Modeler~\cite{stem}. As an extension of the Eclipse development environment, it uses its graphical layout and the general philosophy of user interaction. It requires the user to specify a medium where the simulation is performed with its discretisation level (i.e. whether the node is a municipality or an entire county). The next step is to attach an appropriate solver propagation model and set the starting parameters of the experiment. Once this is done, it is possible to visualise the spread progress. In addition to the extensive user interface, there is an option to run simulations in headless mode, accessible from the terminal. This software also allows for simulating (to a limited extent) the propagation of coexisting phenomena, such as vaccine vs. disease. However, this package is primarily designed for epidemiologists, and all of its functionalities are related to this subject. Therefore, like other tools, it does not support spreading in multilayer or temporal networks. 

\subsection{EpiModel}
EpiModel~\cite{jenness2018epimodel} is an R tool for simulating infectious disease dynamics. It contains deterministic compartmental models, stochastic individual-contact models, and stochastic network models. Network models use the robust statistical methods of exponential-family random graph models (ERGMs). As a standard, it includes SI, SIS, and SIR models. EpiModel does not directly support real networks (it focuses on generated ones) and multiple processes spreading at the same time. It also focuses on epidemic models. Unlike previously described tools, thanks to the application of ERGMs, it partially supports multilayer and temporal networks.

\subsection{Summary}
Based on our review of existing tools, we can conclude that, to the best of our knowledge, none of them supports multiple processes at the same time (like information and disinformation, virus and information about the virus, or multiple diseases), and most of them do not support multilayer and temporal networks. That is why we will focus on these functionalities in the section~\ref{sec:examples}, where we present a few examples of experiments using the Network Diffusion library. 

\section{Feature Overview}
\label{sec:features}

After a thorough review of the available 
Conclusions from Sec.~\ref{sec:tools} were a principal motivation for investing time and human resources in developing the Network Diffusion package. In formulating its design, we operated under the following four premises:
\begin{enumerate}
    \item compatibility with other tools commonly used in the domain of data science,
    \item development of a tool in the form of a framework as opposed to a library comprising loosely connected code fragments,
    \item supporting both multilayer and temporal networks,
    \item supporting spreading models with discrete states.
\end{enumerate}

By releasing a Python-based tool, we naturally facilitated access to the entire ecosystem associated with that language. Primarily, we maintained compatibility with the widely-used graph processing package, NetworkX, which is the backbone of Network Diffusion. It is good to note, that some parts of the library are implemented in C --- that allows to speed up computations with high computational cost.

The framework-oriented design approach enabled the straightforward extension of the tool with new spreading models. That is important, especially in research on social influence, misinformation, or viral marketing, where nuanced phenomena require modelling tailored to their unique nature. Furthermore, this approach allows for incremental project advancement (e.g., as a side effect of in-depth research).

As outlined in Sec,~\ref{sec:tools}, only a few tools for simulating spreading phenomena operate on temporal or multilayer networks, despite these graph models offering a more accurate reflection of reality. This fundamental disparity led us to present our internal research environment as a packaged solution. It is good to note that during the implementation of network models, the library was enriched with centrality methods tailored to these specific structures. These measures can be efficiently used to select seed nodes for diffusion processes.

The assumption regarding the discrete nature of dissemination models undoubtedly constitutes a tool limitation. However, the authors have yet to encounter the need to use a different class of models in their research so far. Therefore, this aspect, being unobjectionable, has been omitted.

While describing the functionality, it's worth mentioning the experiment scheme facilitated by this library. The process entails three fundamental steps: defining the propagation model, specifying the network upon which the model will be executed, and determining the simulation parameters (e.g., number of steps). With these three components in place, experimenting becomes feasible, yielding precise data concerning the network's state at each simulation step. Sec.~\ref{subsec:sir_ua} demonstrates this procedure using the appropriate modules within the package, employing a selected research problem.

\section{Experiments with Network Diffusion}
\label{sec:examples}

This chapter presents four challenges from the domain of spreading models, which are tackled using the Network Diffusion library. Please note that all experiments described below are intended to meet two goals. Firstly, to present various functionalities of the Network Diffusion library. Secondly, despite their compactness, they are big enough to draw particular conclusions regarding issues like phenomena spreading or network modelling. Regarding the above, we do not provide a comparative analysis of the methods used --- this paper is not a place for that, and dwelling on such details could disturb the article's main message.

All experiments were conducted in an environment based on Python 3.10 with Network Diffusion 0.13.0. Though simulations were executed parallelly, we used three computers for that. The first one, which served in problems from Sec.~\ref{subsec:sir_ua}~\ref{subsec:icm_mds}, had installed macOS 13 with M2 Pro CPU (arm64 architecture) and 32 GB RAM. The second one was used in experiments from Sec.~\ref{subsec:ltm_cogsnet} and had installed Ubuntu 20.04 with Intel Core i7-6500U (x86\_64 architecture) CPU and 16 GB RAM. Simulations presented in Sec.~\ref{subsec:nem} were conducted on a workstation with Windows 10 Pro (22H2), whose hardware consisted of Intel Core i7-6820HQ (x86\_64 architecture) CPU and 32 GB RAM. It is good to note that we provide the source code enabling the reproduction of all experiments. It can be found in the repository available on GitHub (\href{https://github.com/anty-filidor/bdma-experiments/tree/08df5793bb69f4792a92d05644b431b3d96516a3}{https://github.com/anty-filidor/bdma-experiments}) with the instructions on how to run it.

\subsection{Problem: simultaneous spreading of disease and awareness of the threat}
\label{subsec:sir_ua}

In this chapter, we demonstrate how to define a custom spreading model using pre-existing library interfaces and how to conduct an experiment. Two processes (disease and awareness of its existence) interacting with each other will serve as examples. First, the SIR (Suspected-Infected-Removed) epidemiological model with parameters specific to COVID-19~\cite{Toda2020Mar}, namely the infection rate coefficient ($\alpha$) and recovery rate coefficient ($\beta$). According to the simulated scenario, the second phenomenon, i.e., awareness of the disease UA (Unaware-Aware), is supposed to reduce the infection rate among agents who are conscious of the contagion. Finally, three cases are considered: minimising social interactions (e.g. lockdowns), wearing masks, and the absence of infection countermeasures. With the experiment, we will determine the effectiveness of the two measures mentioned above in combating the rise of the epidemic.

\subsubsection{Problem formulation}

\begin{figure}[ht]
    \centering
    \includegraphics[scale=0.65]{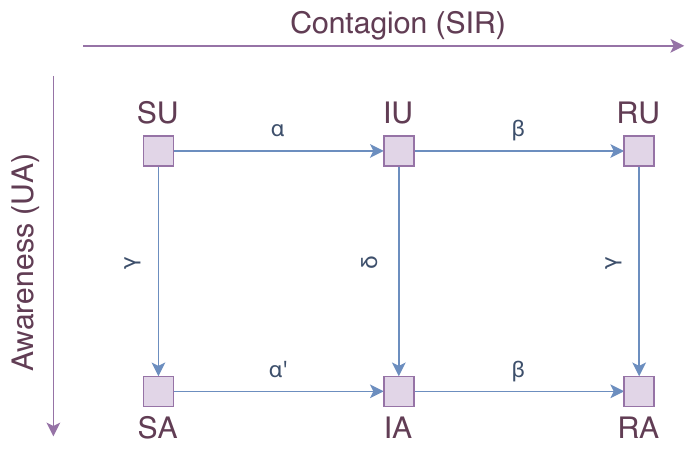}
    \caption{Graph of states and transitions between them for the spreading model of two processes: contagion (SIR) and awareness of its existence (UA). Each state is represented by two letters indicating the state of both processes, e.g., IU indicates that the node is Infected (\textit{I}) with the disease and Unaware (\textit{U}) of its existence. The symbols on the arrows indicates the transition probability from one state to another (for values please see Tab.~\ref{tab:sir_ua}).}
    \label{fig:sir_ua_processes}
\end{figure}

That scenario can be depicted as a state graph with transitions between states, as shown in Fig.~\ref{fig:sir_ua_processes}. The horizontal transitions occur within the disease context, while the vertical transitions occur within the awareness. For instance, the transition $SA \xrightarrow{} IA$ signifies that an agent's state change from suspected and aware to infected and aware takes place with a probability of $\alpha'$. 

It is important to note that the recovery rate does not depend on the awareness of the disease, i.e. $IU \xrightarrow{} RU$ and $IA \xrightarrow{} RA$ both have the same weight denoted as $\beta$. On the other hand, concerning the infection rate, we have assumed that $\alpha'$ will be a reduced value of $\alpha$ by a $\lambda$ factor appropriate to the simulated scenario (see Tab.~\ref{tab:sir_ua}). Thus, for lockdown, we will set infection risk reduction to 90\% ($\lambda = 0.1$)~\cite{wkatroba2023influence}, for wearing masks or 1m social distancing to 65\% ($\lambda = 0.35$)~\cite{Chu2020Jun, McGrail2020Jul}, and for no countermeasures applied no risk reduction will take place, hence $\lambda = 1$. 

In the case of the UA process, we have assumed that agents initially become aware of the disease with a constant and low probability $\gamma$. However, if they get infected, this value increases proportionally to the presence of symptomatic cases in the population observed during the COVID-19 pandemic ($70\%$ of people having COVID-19 had symptoms of the disease)~\cite{Alene2021Mar}. Transition weights between states are also included in Tab.~\ref{tab:sir_ua}.

\begin{table}[ht]
    \centering
    \caption{Transition probabilities (or weights) and their interpretation in the SIR-UA model.}
    \begin{tabular}{c|c|p{3.3cm}}
    Symbol & Formula / Value & Description \\ \hline
    $\alpha$ & $0.19$ & probability of infection for unaware agents \\ \hline
    $\alpha'$ & \makecell{$\lambda\alpha;$ \\ $\lambda \in \{ 0.1, 0.35, 1\} $}  & probability of infection for aware agents \\ \hline
    $\beta$ & $0.10$ & probability of recovery \\ \hline
    $\gamma$ & $0.01$ & probability of awareness for uninfected agents \\ \hline
    $\delta$& $\gamma + 1 - 0.3$ & probability of awareness for infected agents \\
    \end{tabular}
    \label{tab:sir_ua}
\end{table}

Another crucial aspect is the environment where the processes propagate. In this case, we assume propagation occurs in a multilayer network with two layers. In one layer (e.g. a communication network), awareness of the disease spreads, while in the second layer (e.g. a network of physical contacts), the disease itself. Agents can change their state based on interactions with their neighbours, with the probability of adopting the neighbour's state corresponding to the transition weight according to Fig.~\ref{fig:sir_ua_processes} and Tab.~\ref{tab:sir_ua}. An exception is the transitions $IU \xrightarrow{} RU$, $IA \xrightarrow{} RA$, for which no interaction with neighbouring agents is needed for the transitions to occur.

\subsubsection{Experiment setup}

In the repository attached to the article, within the script \textit{utils/models.py}, we have included the source code for the \lstinline[style=py]{SIR_UAModel} class, which represents the implementation of the problem discussed in this chapter. It also demonstrates how to extend the Network Diffusion library with custom spreading processes. To create a specific model, one needs to extend the base class (which is \lstinline[style=py]{network_diffusion.models.BaseModel}) and concrete the six methods accordingly.

The constructor (\lstinline[style=py]{SIR_UAModel.__init__} accepts nine parameters, of which six are the transition weights, and the remaining two are the initial percentage of infected agents and aware agents. The class utilises the transition graph introduced in our previous work~\cite{czuba2022networkdiffusion}, which helps in determining the potential state change of the evaluated agent during the simulation.

The following method determines the initial state of the network before the start of the simulation (\lstinline[style=py]{SIR_UAModel.determine_initial_states}). In this function, seeds are chosen, that is, the infected agents and aware agents from which the process will propagate. The method returns a list of \lstinline[style=py]{NetworkUpdateBuffer} structures, which contain the necessary information for updating the network state (specifically, the agent's ID and states in each layer). In this problem, we aim to determine the impact of $\lambda$ on the number of infections. Therefore, seed nodes are decided to be chosen randomly --- that approach is the most transparent in presented circumstances. It is good to add that we will select initially infected and aware actors independently, with budgets equal to 5\%.

The function \lstinline[style=py]{SIR_UAModel.agent_evaluation_step} is called during each simulation step for each agent individually. It takes the agent's ID, the ID of the layer for which the evaluation takes place, and the network itself. This function determines the state the agent will adopt in the next step of the experiment. That is done by reading the set of possible states the agent can fall in and then iteratively attempting to adopt the neighbour's state with the probability as depicted in Fig.~\ref{fig:sir_ua_processes}. The output of this function is the new state of an agent in the given network layer.

A function \lstinline[style=py]{SIR_UAModel.network_evaluation_step} is responsible for evaluating the entire network in a single experiment step. In this case, it involves iterating over each layer and each agent within that layer to determine its new state according to the method (\lstinline[style=py]{SIR_UAModel.agent_evaluation_step}). The output is a list of \lstinline[style=py]{NetworkUpdateBuffer} structures.

The last two methods are needed to generate the experiment report while using this model. These are \lstinline[style=py]{__str__}, which provides a description of the model and \lstinline[style=py]{get_allowed_states}, which returns the textual form of the graph presented in Fig.~\ref{fig:sir_ua_processes}.

A model prepared this way can be executed on a given two-layer network with the \lstinline[style=py]{network_diffusion.Simulator} class \footnote{In the initial version of the package, it was called \lstinline[style=py]{network_diffusion.MultiSpreading}}. By using the \lstinline[style=py]{Simulator.perform_propagation} method, we conduct a simulation that conceptually consists of eight steps\footnote{Simulation algorithm is described in a simplified form, and simulations in temporal networks are not covered by the pseudocode} --- see Alg.~\ref{algo:preform_propagation}.

\begin{algorithm*}
\caption{Simplified simulation procedure in Network Diffusion library}
    \begin{algorithmic}[1]
 \Procedure{\lstinline[style=py]{perform_propagation}}{\lstinline[style=py]{network, model, epochs}}
     \State \lstinline[style=py]{states_0} $\gets$ \lstinline[style=py]{model.determine_initial_states()}
     \State \lstinline[style=py]{model.update_network(states_0)} \Comment{Predefined function in the \texttt{BaseModel} class}
     \For{\lstinline[style=py]{e} in \lstinline[style=py]{[1, ..., epochs]}}
  \State \quad \quad \lstinline[style=py]{states_e} $\gets$ \lstinline[style=py]{model.network_evaluation_step(network)}
  \State \quad \quad \lstinline[style=py]{model.update_network(network, states_e)}
     \EndFor
     \State \lstinline[style=py]{logs} $\gets$ generate logs from experiment \\
     \Return \lstinline[style=py]{logs}
 \EndProcedure
    \end{algorithmic}
    \label{algo:preform_propagation}
\end{algorithm*}

\subsubsection{Results \& discussion}

\begin{figure}[ht]
    \centering
    \includegraphics[scale=0.35]{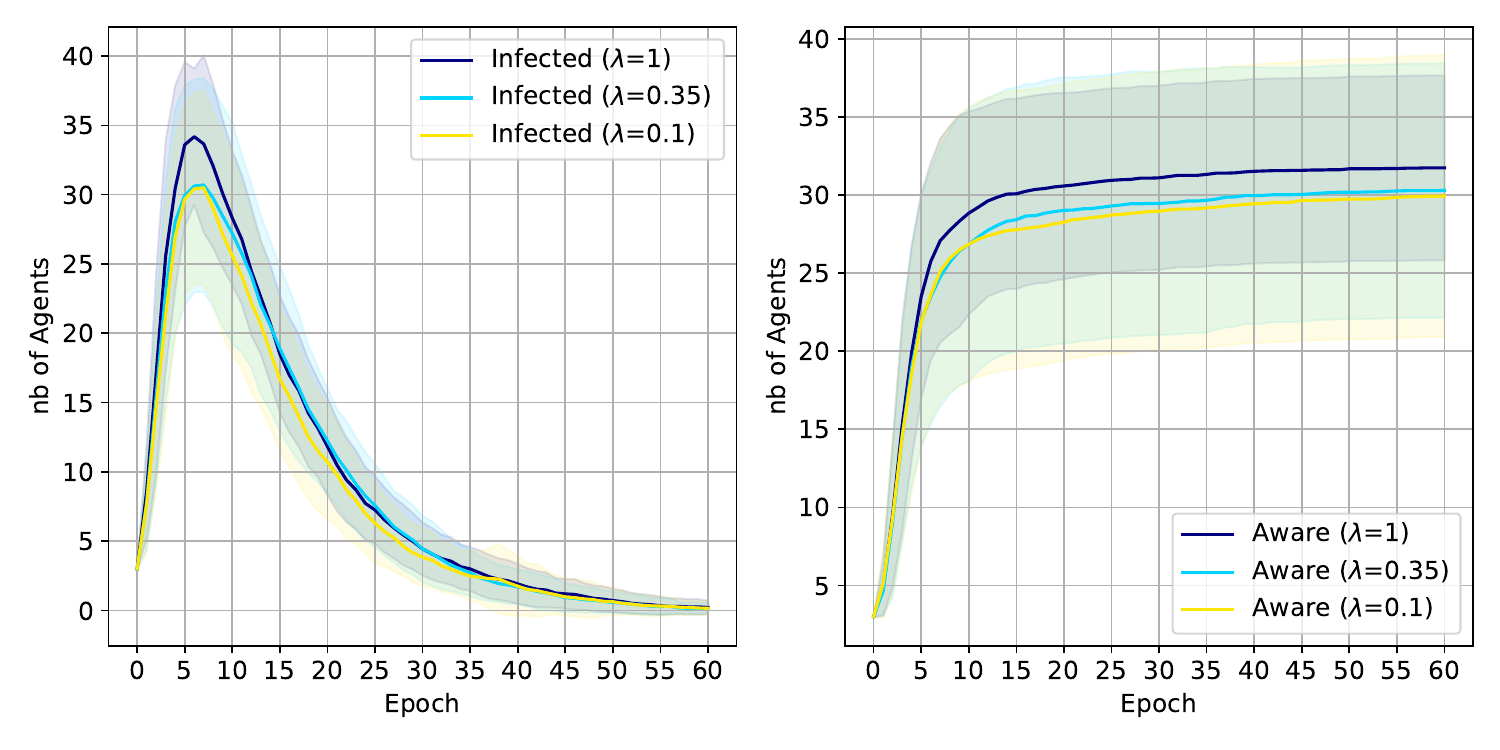}
    \caption{Infection and Awareness curves for spreading of SIR-UA model within \textit{aucs-2} network in three different epidemic regimes where we can observe how different prevention measures lockdown (infection risk reduction by 90\% -- $\lambda = 0.1$), wearing masks or 1m social distancing (risk reduction by 65\% -- $\lambda = 0.35$), and no measures (no risk reduction -- $\lambda = 1$) affect the number of infected and aware individuals (agents) in the network. }
    \label{fig:sir_ua_aucs}
\end{figure}

\begin{figure}[ht]
    \centering
    \includegraphics[scale=0.35]{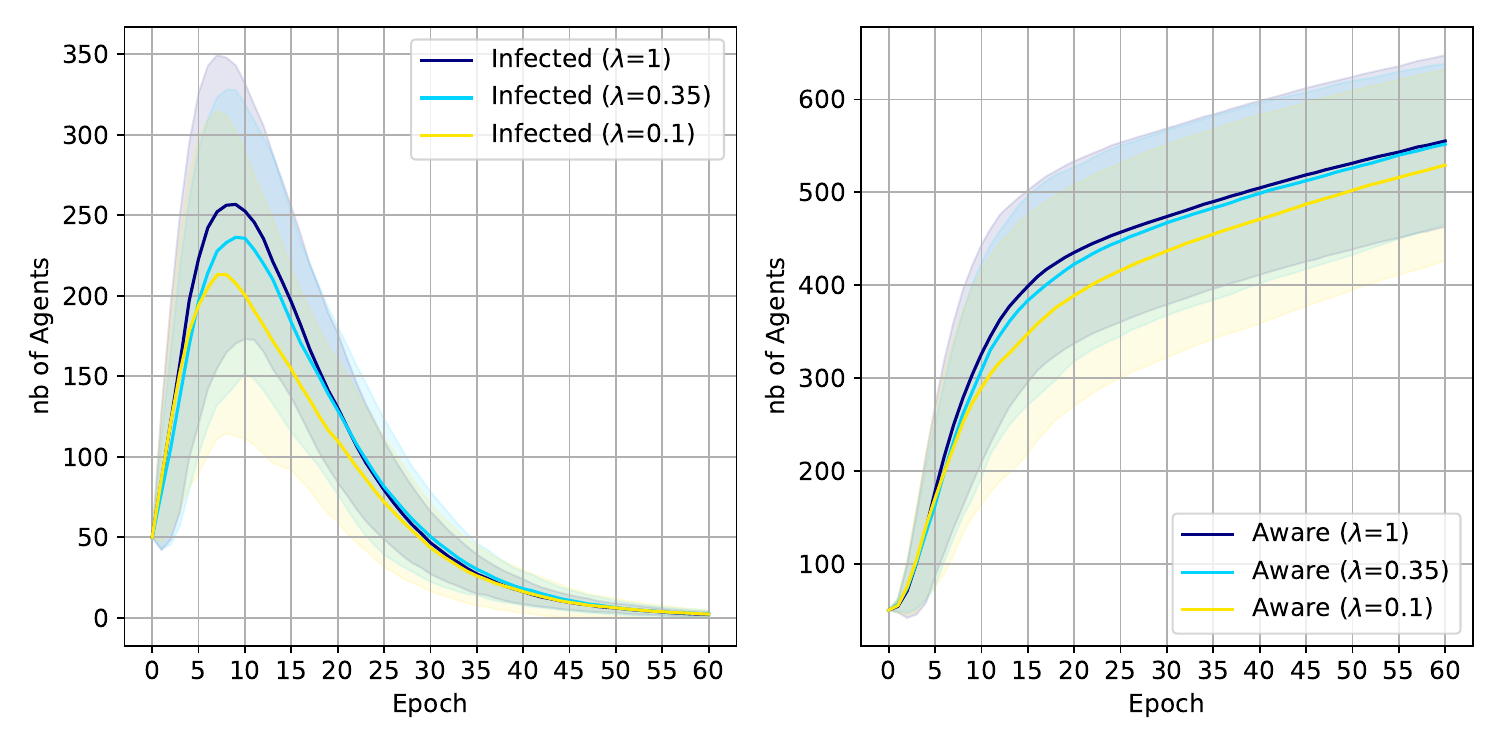}
    \caption{Infection and Awareness curves for spreading of SIR-UA model within \textit{sf-2} network in three different epidemic regimes where we can observe how different prevention measures lockdown ($\lambda = 0.1$), wearing masks or 1m social distancing ($\lambda = 0.35$), and no measures ($\lambda = 1$) affect the number of infected and aware individuals (agents) in the network.}
    \label{fig:sir_ua_sf}
\end{figure}

\begin{figure}[ht]
    \centering
    \includegraphics[scale=0.35]{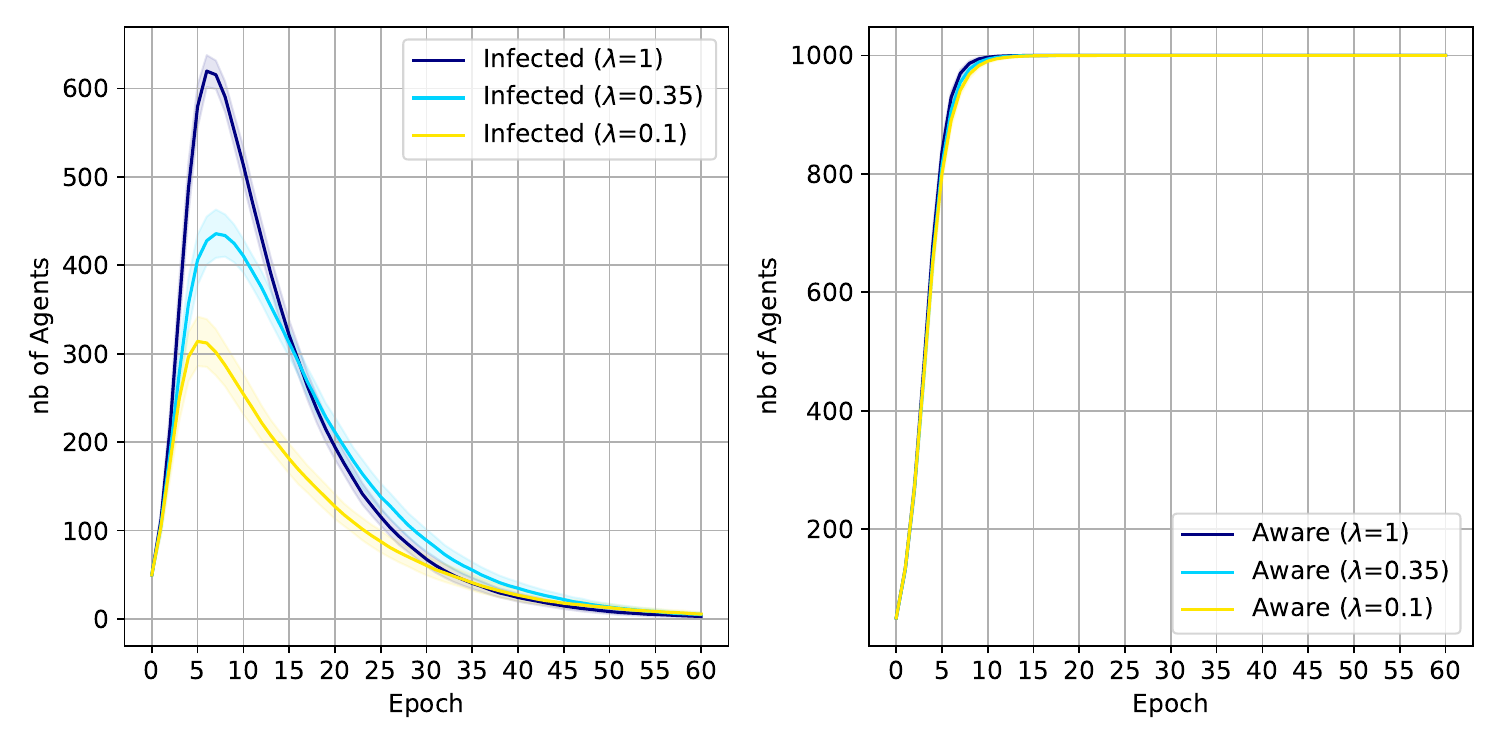}
    \caption{Infection and Awareness curves for spreading of SIR-UA model within \textit{er-2} network in three different epidemic regimes where we can observe how different prevention measures lockdown ($\lambda = 0.1$), wearing masks or 1m social distancing ($\lambda = 0.35$), and no measures ($\lambda = 1$) affect the number of infected and aware individuals (agents) in the network.}
    \label{fig:sir_ua_er}
\end{figure}

As part of the experiment, we utilised three types of networks: AUCS (a real graph obtained from interactions between employees of \textbf{A}arhus \textbf{U}niversity, Department of \textbf{C}omputer \textbf{S}cience), Scale-Free, and Erdős–Rényi, with properties listed in Tab.~\ref{tab:networks_eda}. For each graph, the simulation was repeated 50 times, and the obtained results were averaged. Fig.~{\ref{fig:sir_ua_aucs}--\ref{fig:sir_ua_er}} depicts the results in the form of infection and awareness curves within the population for each evaluated $\lambda$ value. It can be observed that limiting the infection rate makes sense for every network type. Another conclusion is that the network structure clearly influences the course of the processes. The highest percentage of infected agents at a given time was observed in the random network, accompanied by a rapid saturation of the UA process. We can observe that the propagation dynamics were significantly lower for the Scale-Free graph and AUCS network, where degree distribution is not as uniform as for the Erdős–Rényi model. Consequently, the number of infected agents was lower as well.

\subsection{Problem: Linear Threshold Model in temporal networks}
\label{subsec:ltm_cogsnet}

\subsubsection{Problem formulation}
The models of social influence consider different scenarios of the spread. In a binary-LTM that has been introduced in~\cite{granovetter1978threshold}, the adoption of a belief occurs in a situation where a certain fraction of the neighbours in an ego network of a node shares a new belief. In this case, contrary to the stochastic-ICM, we can consider the model deterministic and --- given that only unidirectional change is possible --- the only stop condition of the model is when no future activations of nodes are possible.

Typically, the LTM model is used in a static network scenario where the set of nodes and edges is fixed. However, given that this approach is simplistic compared to real-world networks~\cite{holme2012temporal}, it is worth considering the relaxation of the time constraint and moving the problem of seed selection to time-varying networks~\cite{michalski2014seed,michalski2015maximizing}. Yet, the change of the setting to temporal networks also requires an adequate choice of a temporal network model, which can be considered as a problem \textit{per se}. As we already mentioned, for the static scenario the stop condition is when no further activation of nodes is possible. Yet it is worth underlining that for the dynamic setting the activations can still be possible due to the network dynamics, so here the process can evolve for longer.

The goal of this experiment is to answer the question of how the network dynamics change the final spread of the LTM compared to the static scenario. Conceptually one can think of such an experiment in a way that the answer to this question will allow for understanding whether there is a speed-up or slow-down of diffusion dynamics when the network evolves over time~\cite{scholtes2014causality}.

\subsubsection{Experiment setup}
In the experiment on the spread of influence following the LTM model, the CogSNet model will be used~\cite{michalski2021social}. The properties of this model -- contrary to snapshot-based approaches -- allow for continuous modelling of temporal networks. Here, each event that happens at a time point either creates an edge between nodes interacting with each other or amplifies it. Next, as time passes, the weights of the edges decay over time and in case of no forthcoming events lead to the vanishing of edges. Otherwise, if events occur, they again lead to an increase in weight. The decaying function is typically a power or exponential function. This approach, described in more detail in~\cite{michalski2021social} allows for fine-graded modelling contrary to incremental or snapshot-based temporal network models. However, as there is no continuous variant of the LTM model (the closest to it being introduced in~\cite{karimi2013threshold}), to match the temporal network with the LTM, we will be matching the iterations of the spreading model with intervals of time in the CogSNet network. The stop condition for the LTM was when no further activations were possible and -- for the temporal case -- the process was run for no later than the last edge appeared.

To investigate how much the temporal network speeds up or slows down the spread of influence, the experimental setting was as follows. For experiments, we used the dataset on email exchange in manufacturing company~\cite{nurek2020combining}. For the LTM, we focused on evaluating the seeding budget $\gamma$ and the threshold $\mu$ for node activations. For the seed selection strategy, a random seed selection has been applied, thus the experiments were repeated 100 times. As for the CogSNet model parameters, based on the work~\cite{michalski2021social}, we used the exponential forgetting function, edge lifetime of seven days, edge removal threshold $\theta=0.1$, and forgetting function parameter $\lambda=0.8$, with the units of hours. This set of parameters of the model in the aforementioned work resulted in an adequate match of human memory imprints and ground truth data. The parameters of the experiments alongside their explanation are shown in Tab.~\ref{tab:ltm-parameters}.

\begin{table}[ht]
    \centering
    \caption{Parameters used in the LTM in temporal networks experiment.}
    \begin{tabular}{c|c|p{3.3cm}}
    Symbol & Formula / Value & Description \\ \hline
    $\theta$ & $0.1$ & edge removal threshold \\ \hline
    $\lambda$ & $0.8$ & forgetting function parameter \\ \hline
    $\mu$ & \makecell{$\mu \in \{0.05, 0.1,$  \\ $0.15, 0.2, 0.25, 0.5\}$} & threshold of influence for LTM model \\ \hline
    $\gamma$ &  $\gamma \in \{ 1, 5, 15, 25, 50 \} $ & seeding budget (in \%) \\
    \end{tabular}
    \label{tab:ltm-parameters}
\end{table}
\subsubsection{Results \& discussion}

\begin{figure}[ht]
    \centering
    \includegraphics[scale=0.3]{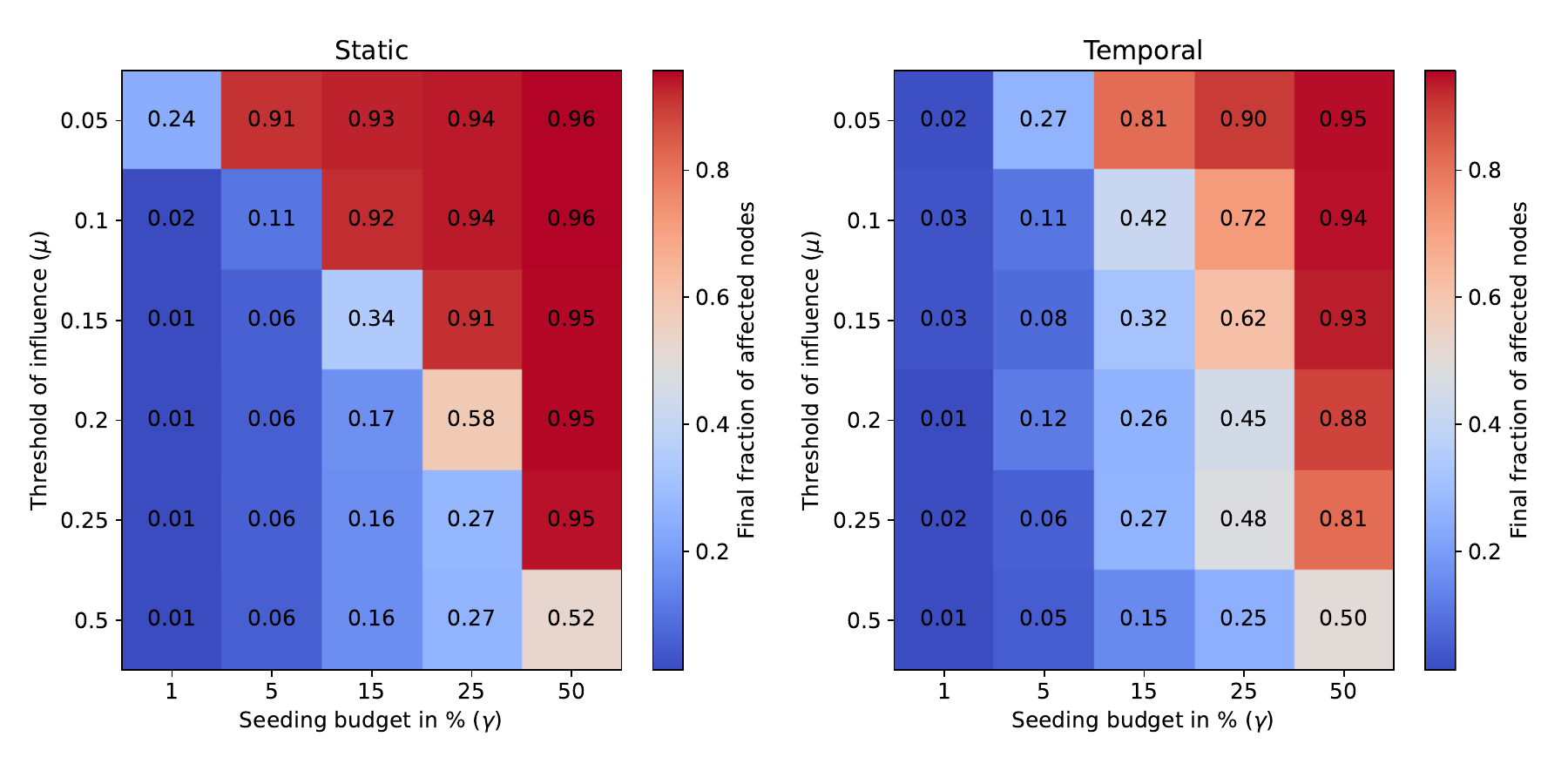}
    \caption{The comparison of a final spread as a percentage of activated nodes for the LTM of social influence for two settings: the temporal network based on the CogSNet model and the aggregated static network for the e-mail exchange data in a manufacturing company. The evaluated parameters were the seeding budget ($\gamma$) and the LTM threshold ($\mu$).}
    \label{fig:ltm_cogsnet_vs_static}
\end{figure}

The results of the experiment depicted in Fig.~\ref{fig:ltm_cogsnet_vs_static} clearly indicate that for the static networks --- compared to the temporal one --- for combinations of high seeding budget and low threshold, a static network is easily getting penetrated by the LTM diffusion process reaching almost 100\% nodes existing in the network. This is not that surprising since in the static network that collapses all the links, there are more paths to reaching other nodes. Contrary to that, in the same regime of threshold values and percentage of initially activated nodes, the spread slows down in the temporal network built upon the CogSNet model. That can be caused by the absence of some potential links that did not appear in the later stage of the network evolution and that could not have been used for activation.

Yet, there is one interesting observation that one can make when looking at the combination of the LTM model parameters that are more challenging for the diffusion process (the central part of Fig.~\ref{fig:ltm_cogsnet_vs_static}). Here, the CogSNet-modelled temporal network led to a higher number of activations. It is due to the fact that there were fewer links in the ego networks of activated nodes that had to be activated to change the state of the node. This means that the lower density of the network results in a higher number of activations in this regime. Finally, both approaches perform equally badly in the most difficult scenario (left side of Fig.\ref{fig:ltm_cogsnet_vs_static}).

\subsection{Problem: multilayer Independent Cascade Model initialised with Minimum Dominating Set}\label{subsec:icm_mds}

\subsubsection{Problem formulation}

The influence maximisation problem is well-known since the work of~\cite{kempe2003maximizing} and adopts various forms~\cite{Shashank2022InfluenceMaximization}. Studies have shown that the ultimate diffusion effectiveness depends not only on the spreading process itself and its parameters but also significantly on the network structure~\cite{albert2022statistical, boccaletti2006complex, liu2009alternative}. In general, ranking methods for selecting seed nodes can be categorised into three groups: local, semi-local, and global~\cite{lu2016vital}, differing in the scope of the network needed to analyse to determine the metric's value for a given node. Each of these classes varies in the trade-off between accuracy and computational complexity. Therefore, research teams often have to accept lower accuracy in identifying super-spreaders to obtain results within a reasonable time on large networks. Occasionally, one can leverage additional features to select the seed set, e.g. node labels, which are often available when analysing data derived from social networks, as exemplified in~\cite{forouzandeh2018new}. However, some studies have shown that it is possible to determine (by analysing the network's structure) whether computationally expensive methods can surpass basic algorithms such as degree centrality~\cite{berahmand2018effect}.

Maximising influence problem has long transcended the classical defined graphs as a set of nodes and edges. This experiment will focus on multilayer networks~\cite{dickison2016multilayer}. By utilising the Network Diffusion framework, we will conduct a study to determine (in a minimal extent) whether network control methods are suitable for selecting initial nodes for propagation in graphs of this type. In order to build new seed selection strategies for the spreading processes in networks, we adopted the concept of driver nodes from the control theory. To control a system, one needs first to identify the minimum set of nodes (called driver nodes) that, when driven by distinct signals, can offer full control over the network. We argue that network spreading processes can be viewed as a soft version of control, where control diffuses with a given probability of following certain rule(s). Thus, the intuition is that using driver nodes as seed nodes can result in a more efficient spread over the network, and indeed, it has been successfully employed for classical graphs~\cite{sadaf2023maximising}.

Networks with different characteristics feature different numbers of driver nodes~\cite{sadaf2023effects}. Once the driver nodes are identified, various centrality measures can be used to rank the driver nodes and use this final ranking list in the seed selection process. Experiments show that when driver nodes are used as the basis to build the ranking list, the spreading process is more effective and efficient when compared with the benchmarks. The first study that employed control-theoretic approaches for seed selection utilised the Minimum Dominating Set (MDS) to identify a candidate set of driver nodes~\cite{sadaf2023maximising}. This method has been incorporated into the Network Diffusion framework and extended to work with multilayer networks.

MDS is an optimisation approach that determines the minimum dominating set of nodes in undirected networks~\cite{nacher2012dominating}. MDS is the smallest subset of nodes such that every node of a network either belongs to this subset or is adjacent to at least one node in this set. Formally, a dominating set of a graph $G$ is a subset $D$ of the vertices of $G$ such that every vertex $v$ of $G$ is either in the set $D$ or $v$ has at least one neighbour that is in $D$.

Each driver node (or in the case of multilayer networks: actors) can control its associated nodes (actors) independently, and each non-driver node (actor) is controllable if it is at least adjacent to one driver node (actor). In the MDS method, each node can exert control over its immediate neighbours at the same time, but the control does not propagate further. The driver nodes (actors) are chosen based on a minimal set of nodes (actors) that ensures each node (actor) in the network is either a driver node (actor) itself or directly connected to one. MDS has been applied in various contexts, such as characterising the perturbation of disease genes in the human regulatory network~\cite{wang2015diversified} and identifying control variables in protein interaction networks~\cite{wuchty2014controllability}.

\subsubsection{Experiment setup}
In order to demonstrate the use of driver actors as the seeds in the spreading process, we run an experiment using AUCS~\cite{rossi2015aucs} and Lazega~\cite{snijders2006lazega} multilayer networks and apply an ICM~\cite{kempe2003maximizing}.

In the ICM we start by activating a set of nodes (seeds) which become active. Each active node (in the case of a multilayer network an actor) has one chance to activate each of its neighbours. It succeeds with some probability, known as activation (or propagation) probability. After one iteration, each active node becomes activated and cannot activate anyone any more. At this point, it is good to note that in the case of multilayer networks actors, not nodes, are the subject of the diffusion (i.e. we care whether the actor is active or not, not nodes that just represent the actor on the particular layers). Therefore, one needs to define how to aggregate impulses from layers of the network influencing the actor. We adopted here the approach proposed originally for LTM extended for multiplex networks (described in~\cite{zhong2022mltm}) called "protocol functions" (or strategies). In this study, we used two extreme ones: $OR$, i.e. an actor gets activated when it gets activated on at least one layer and $AND$, i.e. an actor gets activated when it gets activated on all layers it is represented in. It is good to add that between them, there is an entire spectrum of all possible functions that one can invent.

For the first experiment, we used an activation probability of 10\% and AND strategy. We tested two different values of the seeding budget, i.e. 5\% and 10\%. To show how the MDS method performs in the context, we compared it with seed selection strategies based on Degree, Betweenness~\cite{brandes2001Betweenness}, Closeness~\cite{freeman1978closeness}, Vote Rank~\cite{zhang2016voterank}, and Berahmand's centrality~\cite{berahmand2018centrality}. They were adapted to work with multilayer networks as follows: Closeness, Betweenness and Berahmand's centrality were computed separately in each layer of the graph, and then final values for actors were obtained by averaging scores from each layer. We modified Vote Rank by replacing the Degree of an actor with Neighbourhood size (both comprehended as in~\cite{brodka2021sequential}). Finally, MDS was obtained as a union of minimal dominating sets calculated on each layer of the network. We transformed it into a ranking list by sorting according to the Degree of actors. Averaged results depicting the efficiency of all evaluated methods can be found in Tab.~\ref{tab:dd_AUCS} and Tab.~\ref{tab:dd_Lazega}. 

In order to show the dynamics of the spreading process triggered by actors belonging to MDS (sorted by degree centrality), we conducted a second simulation. For that, we utilised the AUCS network and multilayer ICM, with a 10\% seeding budget, 50\% activation probability and OR strategy. The illustration of the obtained results is depicted in Fig.~\ref{fig:spreading}.

\subsubsection{Results \& discussion}
The results for the AND strategy show that the MDS method performs at a similar level as other methods on the two selected datasets regardless of the seeding budget, and it takes a similar number of epochs to reach a stable state where no more new nodes are being activated (Tab.~\ref{tab:dd_AUCS} and Tab.~\ref{tab:dd_Lazega}). However, it should be noted that the parameters of the spreading model are used for illustration purposes only. The optimal ones are shown to depend on the network structural characteristics~\cite{sadaf2023maximising} and can be optimised separately, but it is out of the scope of this study.

\begin{figure*}[t]
\centering
\includegraphics[scale=0.35]{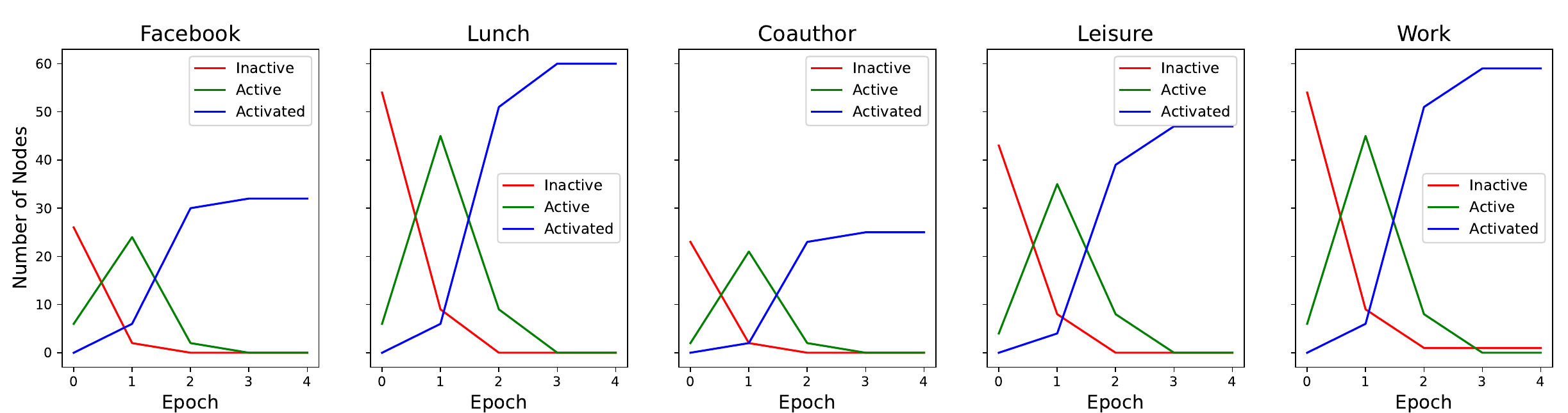}
\caption{The dynamics of ICM spreading in each layer of the AUCS network. The model was executed using MDS (degree centrality) as the seed selection method, with a 10\% seeding budget, 50\% propagation probability and OR strategy. “Active” means that the node is active and has the potential to activate its neighbours; “Activated” denotes the node that is active but does not have the potential to activate its neighbours any more.}
\label{fig:spreading}
\end{figure*}

The results of spreading in the AUCS network using the OR strategy are shown in Fig.~\ref{fig:spreading}. In each of the layers, the number of activated nodes goes through a phase transition between epoch one and two resulting in a sudden increase in the number of activated nodes. This shows that MDS is a good strategy to activate many nodes early in the spread process. Again, used parameters are not optimised and are used for illustrative purposes.

\begin{table*}[ht]
\centering
\caption{Average number of activated nodes for multilayer ICM on AUCS network and average number of epochs needed for the model to reach a stable state when the simulation is stopped. The results for each seed selection method are averaged over 100 runs. The model was executed with 5\% and 10\% seeding budgets, 50\% propagation probability and AND strategy.}
    \begin{tabular}{l|r|r|r|r}
    \hline
    \multirow{2}{*}{\makecell{Seed Selection\\Method}} & \multicolumn{2}{c|}{Seeding Budget 5\%} & \multicolumn{2}{c}{Seeding Budget 10\%} \\
    \cline{2-5} & Avg. Activated Actors ~$\pm$~ Std & Avg. Epochs ~$\pm$~ Std & Avg. Activated Actors ~$\pm$~ Std & Avg. Epochs ~$\pm$~ Std \\ \hline
    Degree   & 11.64~$\pm$~3.17\%  & 2.82~$\pm$~0.77 & 20.90~$\pm$~2.66\%  & 2.79~$\pm$~0.50 \\
    Betweenness     &  9.79~$\pm$~2.51\%  & 2.54~$\pm$~0.68 & 18.11~$\pm$~3.02\%  & 2.76~$\pm$~0.55 \\
    Closeness       & 8.79~$\pm$~1.83\%   & 2.33~$\pm$~0.60 & 16.49~$\pm$~2.43\%  & 2.46~$\pm$~0.54 \\
    VoteRank & 10.98~$\pm$~3.90\%  & 2.76~$\pm$~0.91 & 20.30~$\pm$~3.81\%  & 2.88~$\pm$~0.52 \\
    Berahmand       &  7.72~$\pm$~2.55\%  & 2.10~$\pm$~0.84 & 13.56~$\pm$~1.91\%  & 2.38~$\pm$~0.58 \\
    MDS      & 12.31~$\pm$~3.15\%  & 2.84~$\pm$~0.73 & 19.46~$\pm$~3.84\%  & 2.80~$\pm$~0.51 \\ \hline
    \end{tabular}
    \label{tab:dd_AUCS}
\end{table*}

\begin{table*}[ht]
\centering
\caption{Average number of activated nodes for multilayer ICM on Lazega network and average number of epochs needed for the model to reach a stable state when the simulation is stopped. The results for each seed selection method are averaged over 100 runs.  The model was executed with 5\% and 10\% seeding budgets, 50\% propagation probability, and AND strategy.}
    \begin{tabular}{l|r|r|r|r}
    \hline
    \multirow{2}{*}{\makecell{Seed Selection\\Method}} & \multicolumn{2}{c|}{Seeding Budget 5\%} & \multicolumn{2}{c}{Seeding Budget 10\%} \\
    \cline{2-5} & Avg. Activated Actors ~$\pm$~ Std & Avg. Epochs ~$\pm$~ Std & Avg. Activated Actors ~$\pm$~ Std & Avg. Epochs ~$\pm$~ Std \\ \hline
    Degree  & 79.17~$\pm$~17.74\% & 6.01~$\pm$~0.99 & 88.39~$\pm$~3.52\%  & 4.37~$\pm$~0.52 \\
    Betweenness    & 78.03~$\pm$~15.53\% & 6.02~$\pm$~0.96 & 89.32~$\pm$~3.40\%  & 4.27~$\pm$~0.44 \\
    Closeness      & 78.55~$\pm$~13.07\% & 6.17~$\pm$~0.90 & 89.10~$\pm$~3.35\%  & 4.43~$\pm$~0.57 \\
    VoteRank       & 82.10~$\pm$~13.30\% & 5.84~$\pm$~0.90 & 89.03~$\pm$~3.40\%  & 4.44~$\pm$~0.50 \\
    Berahmand      & 83.10~$\pm$~11.92\% & 5.83~$\pm$~1.02 & 89.13~$\pm$~3.19\%  & 4.31~$\pm$~0.50 \\
    MDS     & 84.90~$\pm$~ 6.04\% & 5.77~$\pm$~0.85       & 89.28~$\pm$~3.28\%  & 4.28~$\pm$~0.45 \\ \hline
    \end{tabular}
    \label{tab:dd_Lazega}
\end{table*}

\subsection{Problem: Network Epistemology Model in temporal networks}
\label{subsec:nem}

\subsubsection{Problem formulation}
Network Epistemology Model (NEM) is an approach to the formalisation of the social learning process, focusing on belief evolution and dissemination in the social network. Previous research on this model explored various aspects, including the impact of community structure~\cite{zollman2007communication}, conformism, and polarization~\cite{oconnor_weatherall_2018_3, oconnor_weatherall_2018, oconnor_weatherall_2018_2, oconnor_weatherall_2020_2} on the effect of the learning process. By adapting the model to a temporal framework, it becomes better equipped to capture the ongoing social network dynamics, aligning it more closely with real-world scenarios~\cite{michalski2022epistemology}.

In its basic configuration, the model assumes the existence of a set of $K$ agents organised within a temporal social network, each facing a bandit decision-making problem. At each iteration of the process, agents are presented with a choice between two available actions. The payoff associated with the first action denoted as $A$, is common knowledge and remains constant at $0.5$. In contrast, the payoff associated with the alternative action denoted as $B$, remains unknown to agents and deviates from the payoff of the action $A$ by a specified value, denoted as $\epsilon$.

Each agent's decision relies on their individual belief level. The belief assigned to each agent is a value within the range of $0$ to $1$ interpreted as the probability that the given agent associates with the proposition that action $B$ yields a superior payoff compared to action $A$.

Subsequent to their decision-making, agents who opt for action $B$ engage in an experimentation phase, collecting evidence related to their actions. They execute the action $B$ a defined $N$ number of times, drawing from a Bernoulli distribution with a probability of success equal to $0.5 + \epsilon$.

Following this experimentation phase, agents update their belief levels using Bayes' rule, calculating posterior probabilities based on their prior beliefs and the evidence they have accumulated. This updating process occurs in a synchronous manner across all agents within the network. The simulation proceeds iteratively through each snapshot of the temporal network.

In the following experiment, we are going to compare the diffusion following NEM between the classical static approach to construct the underlying network topology and the temporal strategy with the CogSNet model~\cite{michalski2021social}. One can notice a similarity with the experiment described in Sec.~\ref{subsec:ltm_cogsnet}. In fact, it exists, but the spreading model used here is much more complex than LTM. Therefore, the results between them are certainly not redundant.

\subsubsection{Experiment setup}
Implementation of the model with an extension allowing to run it in the temporal setting is located in \lstinline[style=py]{TemporalNetworkEpistemologyModel} class. Apart from previously discussed seeding properties, construction of the model instance requires two specific parameters described in the brief model definition, namely $\epsilon$ denoted as \lstinline[style=py]{epsilon}, and $N$ denoted as \lstinline[style=py]{trials_nr}.

As stated above, in the experiment, we used two types of networks: static and dynamic. Both of them were constructed from a dataset on email exchange in manufacturing company~\cite{nurek2020combining}. The number of experiments $N$ that agents perform each iteration is equal to $1$. We test different values of the seeding budget $\gamma$ and values of $\epsilon$, which stands for the difference in the $A$ and $B$ actions expected values. For both cases, we use the random seeding method. The value we compare between these configurations is the final number of actors voting for superior action $B$. The snapshots constituting the temporal network in this experiment were created with one day interval, resulting in 55 long sequences, with one snapshot for each following day of the underlying data. The parameters of NEM can be established based on the expert knowledge related to a specific sociological phenomenon or by a direct measurement of the underlying social learning process characteristic. In this example the fixed parameter $N=1$ corresponds to one decision made daily by each agent during the simulation process. The configuration of CogSNet parameters was based on the work~\cite{michalski2021social} in which it was demonstrated to effectively match the ground truth data on subjects relations. The values of the experiments' outcomes were averaged across 100 iterations to reduce the random effects. Tab.~\ref{tab:nem-parameters} outlines the experiment parameters and their description.

\begin{table}[ht]
    \caption{Parameters used in the Network Epistemology Model in temporal networks experiment.}
    \begin{tabular}{c|c|p{4cm}}
    Symbol & Formula / Value & Description \\ \hline
    $\theta$ & $0.1$ & edge removal threshold \\ \hline
    $\lambda$ & $0.8$ & forgetting function parameter \\ \hline
    \textit{N} & $1$ & number of experiments (trials) performed by an agent per iteration \\ \hline
    $\epsilon$ &  \makecell{$\epsilon \in \{ 0.005, 0.01,$ \\ $0.025, 0.05, 0.1 \}$ } & difference between $A$ and $B$ actions expected values \\ \hline
    $\gamma$ &  $\gamma \in \{ 1, 5, 15, 25, 50 \} $ & seeding budget (in \%) \\
    \end{tabular}
    \label{tab:nem-parameters}
\end{table}

\subsubsection{Results \& discussion}
\begin{figure}[ht]
    \centering
    \includegraphics[scale=0.3]{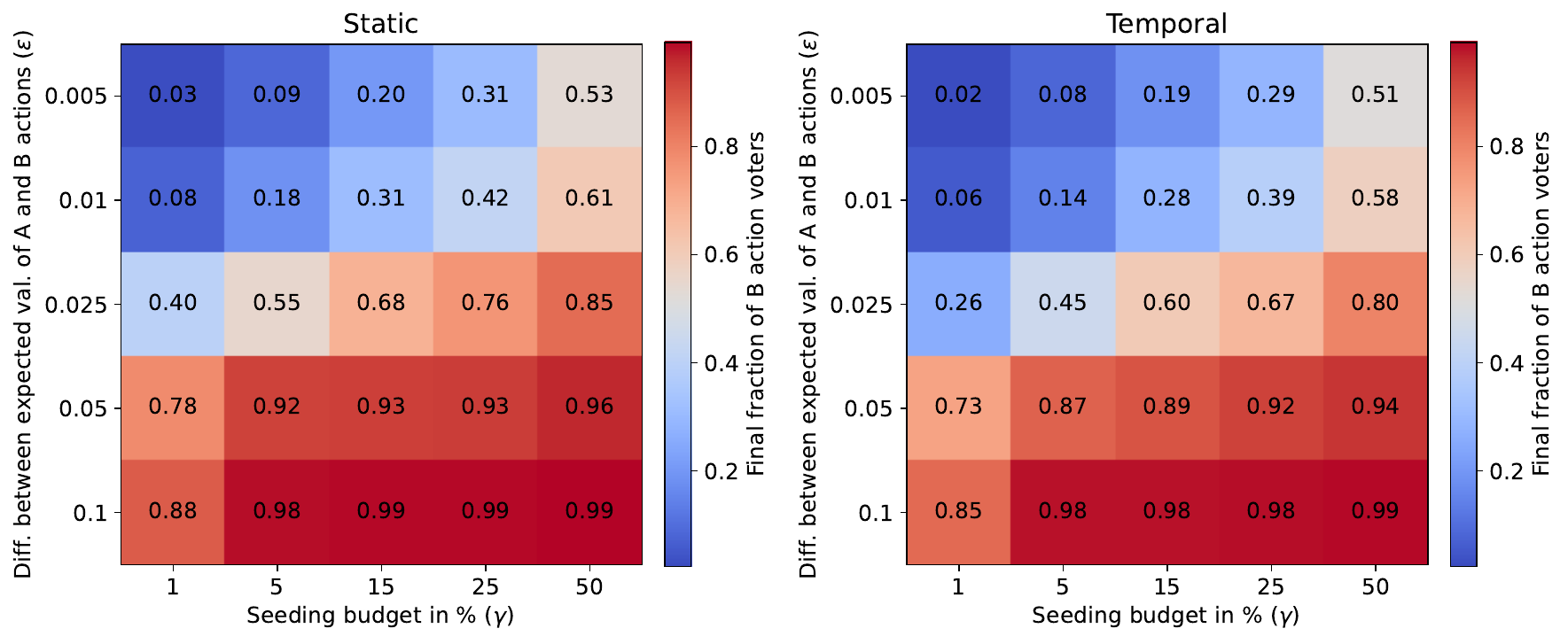}
    \caption{Number of $B$ action voters at the end of the spreading process compared for different levels of problem difficulty and various initial seeding budgets, that in this experiment corresponds to an initial number of $B$ voters. The process outcomes are compared for two different approaches for network topology construction -- a static network and a temporal network built using the CogSNet method.}
    \label{fig:tnem_cogsnet_vs_static}
\end{figure}

The outcomes of the experiment are shown in Fig.~\ref{fig:tnem_cogsnet_vs_static}. The measure of efficiency of the explored configurations is expressed as the ratio of better, $B$ action voters to all individuals. The observed results reveal a subtle difference between the two networks in the effectiveness of $B$ action propagation. On average, in the temporal network, diffusion for each configuration reached a smaller or equal range compared to the static network. The most noticeable differences are in the test configurations, where the $\epsilon$ value was $0.025$, where propagation in the static network covers a few to several percentage points more nodes.

\subsection{Data used in experiments}

As part of the experiments, we utilised several types of networks. Their basic parameters have been listed in Tab.~\ref{tab:networks_eda}. It's worth mentioning that the \textit{aucs-2} network was created with "lunch" and "facebook" layers to simulate the SIR and UA processes, respectively. The \textit{er-2} and \textit{sf-2} graphs were created using the NetworkX~\cite{networkx} library so that the layer responsible for disease transmission was sparser than the one responsible for spreading awareness about the epidemic. Implementation details can be found in the accompanying repository.

\begin{table}[ht]
    \caption{Networks used in experiments with their basic parameters shortlisted.}
    \begin{tabular}{l|r|r|r|r|p{2.2cm}}
    Name & Layers & Actors & Nodes & Edges & Note \\ \hline
    aucs & 5 & 61 & 224 & 620 & \multirow{2}{2.2cm}{AUCS network~\cite{rossi2015aucs}.} \\ 
    aucs-2 & 2 & 60 & 120 & 317 & \\ \hline
    lazega & 3 & 71 & 212 & 1,659 & Lazega Law Firm network~\cite{snijders2006lazega}.\\ \hline
    er-2 & 2 & 1,000 & 2,000 & 27,451 & Erdős–Rényi network~\cite{er-model} \\ \hline
    sf-2 & 2 & 1,000 & 2,000 & 3,357 & Scale-free network~\cite{sf-model} \\ \hline
    manuf & 1 & 151 & 151 & 20,000  & E-mail exchange~\cite{nurek2020combining} \\
    \end{tabular}
    \label{tab:networks_eda}
\end{table}

\section{Limitations \& performance study}
\label{sec:efficiency}

In the last section of the paper, we briefly discuss limitations of the library and complement it with the performance analysis. Apart from the package functionalities described in Sec.~\ref{sec:features}, it is essential to address its constraints critically. They primarily stem from the adopted design assumptions and the relatively small size of the team involved in the implementation.

The most significant drawback of Network Diffusion is also its greatest advantage --- implementation in Python. This results in significantly slower performance compared to if it had been coded in another compiled language. Although a part of the code has been provided in C and is accessible as bindings to Python, we cannot adopt this approach in the entire project due to the small team. Another package limitation is its support solely for discrete spreading processes (see Sec.~\ref{sec:features}). Though the framework was designed more like a set of interfaces to be used in the implementation of custom experiments, it does not include many concrete spreading models that work “out of the box”. Comparing that to NDlib, which consists of dozens of pre-defined models, one can find that state to be a limitation. Finally, it is worth noting the absence of a user interface, which may prove impossible for individuals lacking programming skills, leaving them compelled to resort to tools like NetLogo~\cite{wilensky1999netlogo}.

In order to assess the preformance of the Network Diffusion, we utilised all four models described in the article.The domain chosen to express complexity is the experiment duration as a function of the network size, expressed in the number of actors. We evaluated the following set of parameters:
\begin{itemize}
    \item number of simulation steps --- 200 (regardless of whether a stable state was achieved earlier or not),
    \item evaluation was performed on two-layer multiplex Erdős–Rényi networks with an edge creation probability of 0.1 and the number of actors from range 10 to 1000,
    \item each experiment was repeated 10 times.
\end{itemize}

\begin{figure}[ht]
    \centering
    \includegraphics[scale=0.45]{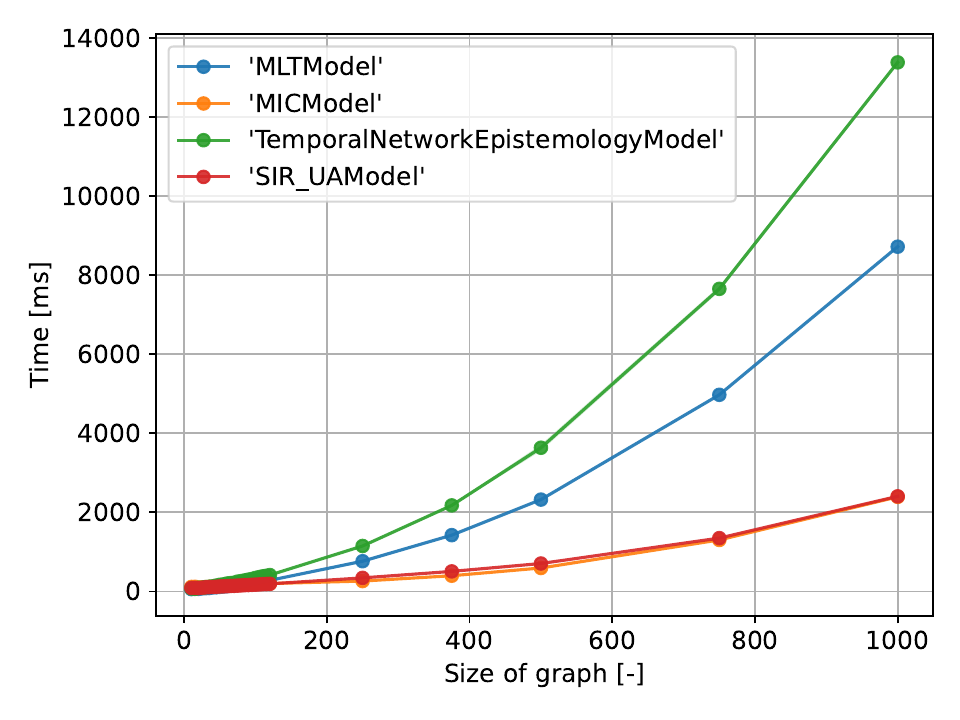}
    \caption{Complexity of spreading models used in experiments; evaluation was performed on two-layer multiplex Erdős–Rényi networks.}
    \label{fig:efficiency}
\end{figure}

The results are presented in Fig.~\ref{fig:efficiency}. As can be seen, the most computationally expensive models were NEM and LTM, which significantly differ from ICM and SIR-UA. It is also worth noting that the computation time is stable --- a standard deviation (a blurry region surrounding curves) is insignificant.

Considering the future of the package, we undoubtedly are going to develop and maintain it. Having defined the framework, it is natural to enhance its contents by new spreading models and functions regarding both temporal and multilayer networks. We are also going to implement machine learning-based methods for the identification of the most influencing actors. Nonetheless, there is another path that is especially interesting to us --- matching the theoretical spreading model to the observed phenomena "in the wild". That will be addressed within the Network Diffusion or at least with its support.  

\section{Conclusions}
\label{sec:conclusions}

Developing and making publicly available comprehensive software frameworks that support replicability and dissemination of research effort is a key and very much required component of scientific enquiry. In order to support the research community, we showed in this paper an extended version of the Network Diffusion framework. Its initial concept was presented in~\cite{czuba2022networkdiffusion} and has now been significantly enriched to encompass both temporal and multilayer networks, as well as various methods and spreading processes that can propagate over different types of graphs. 

The models covered in this study include multilayer networks and temporal networks. Additionally, we adopt the node centrality measures for multilayer networks. Spreading models covered in this paper include SIR-UA, temporal LTM, temporal NEM, and multilayer ICM. Presented spreading models, which have been included in the framework's extension, are accompanied by experiments to show how the newly developed parts of the library can be utilised. They also show the interesting future direction of our research that we would like to investigate in the future fully. 

As discussed above, the Network Diffusion framework is complementary to other software solutions available in the field of data science. We believe that its comprehensiveness and ease of use make it accessible not only for computer scientists, but also for researchers from other domains who are interested in understanding spread in the systems they work on.

\section*{Authors Contribution}
Conceptualization (MCz, PB), Data curation (MCz, YQ, MN, RM), Formal Analysis (All), Funding acquisition (PB, RM), Investigation (All), Methodology (MCz, PB, RM, KM), Project administration (MCz), Resources (All), Software (MCz, DS, MN, MJ, YQ), Supervision (PB), Validation (All), Visualization (MCz, DS, MN, MJ, YQ), Writing – original draft (All), Writing – review \& editing (MCz, PB).

\section*{Acknowledgments}
This research was partially supported by the National Science Centre, Poland, under Grant no.  2022/45/B/ST6/04145 and 2021/41/B/HS6/02798, and the EU under the Horizon Europe, grant no. 101086321 OMINO. Views and opinions expressed are, however, those of the authors only and do not necessarily reflect those of the National Science Centre, EU or the European Research Executive Agency.

\bibliographystyle{IEEEtran}
\bibliography{bibliography}

\vskip -2\baselineskip plus -1fil

\begin{IEEEbiography}
[{\includegraphics[width=1in,height=1.25in,clip,keepaspectratio]{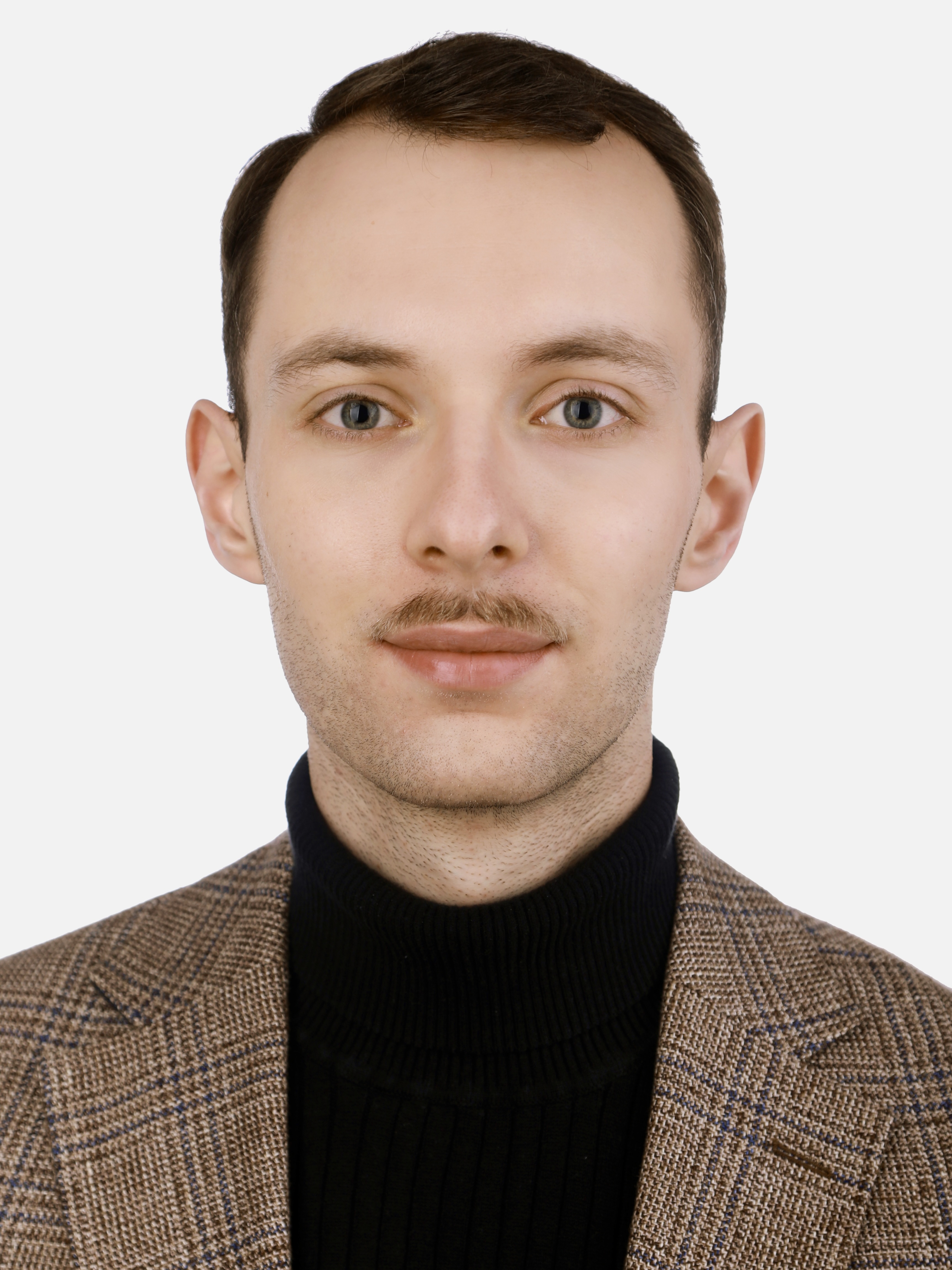}}]{Micha{\l} Czuba}
is a PhD candidate at Wroc{\l}aw University of Science and Technology, in the discipline of Information and Communication Technology with a focus on computational network science, namely problems of influence maximisation and phenomena spreading in multilayer networks. He also has industrial expertise in machine learning, computer vision systems and MLOps engineering in projects concerning the commercialisation of research outcomes.
\end{IEEEbiography}
\vskip -2\baselineskip plus -1fil

\begin{IEEEbiography}
[{\includegraphics[width=1in,height=1.25in,clip,keepaspectratio]{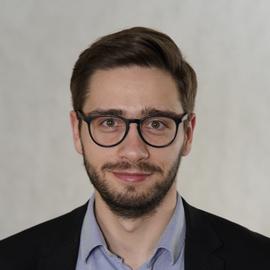}}]{Mateusz Nurek}
is a PhD student at Wroc{\l}aw University of Science and Technology. His research area includes network science and machine learning. Mateusz is primarily interested in using computational intelligence to study social aspects; therefore, his current research focuses on problems such as the classification of human relationships or predicting personality traits based on communication patterns.
\end{IEEEbiography}
\vskip -2\baselineskip plus -1fil

\begin{IEEEbiography}
[{\includegraphics[width=1in,height=1.25in,clip,keepaspectratio]{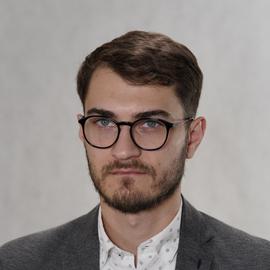}}]{Damian Serwata}
is a PhD student at Wroc{\l}aw University of Science and Technology. His research focus is concentrated on leveraging network science models and tools to study complex social systems and their adaptive abilities. Currently, he is devoted to research on the social learning process, with regards to the investigation of different approaches to represent social structures, bringing the models closer to reality.
\end{IEEEbiography}
\vskip -2\baselineskip plus -1fil

\begin{IEEEbiography}
[{\includegraphics[width=1in,height=1.25in,clip,keepaspectratio]{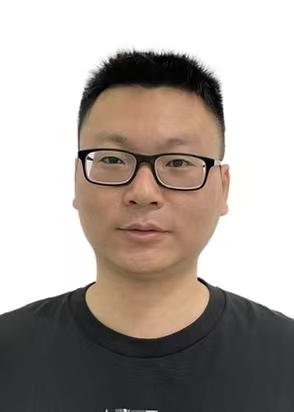}}]{Yu-Xuan Qiu}
presently serves as a PostDoc Research Assistant at the University of Technology Sydney. He earned his BE and ME degrees in Computer Science from Shenzhen University, China, in 2015 and 2018, respectively, and obtained his PhD from the University of Technology Sydney in August 2023. His primary research interest is graph data mining and management.
\end{IEEEbiography}
\vskip -2\baselineskip plus -1fil

\begin{IEEEbiography}[{\includegraphics[width=1in,height=1.25in,clip,keepaspectratio]{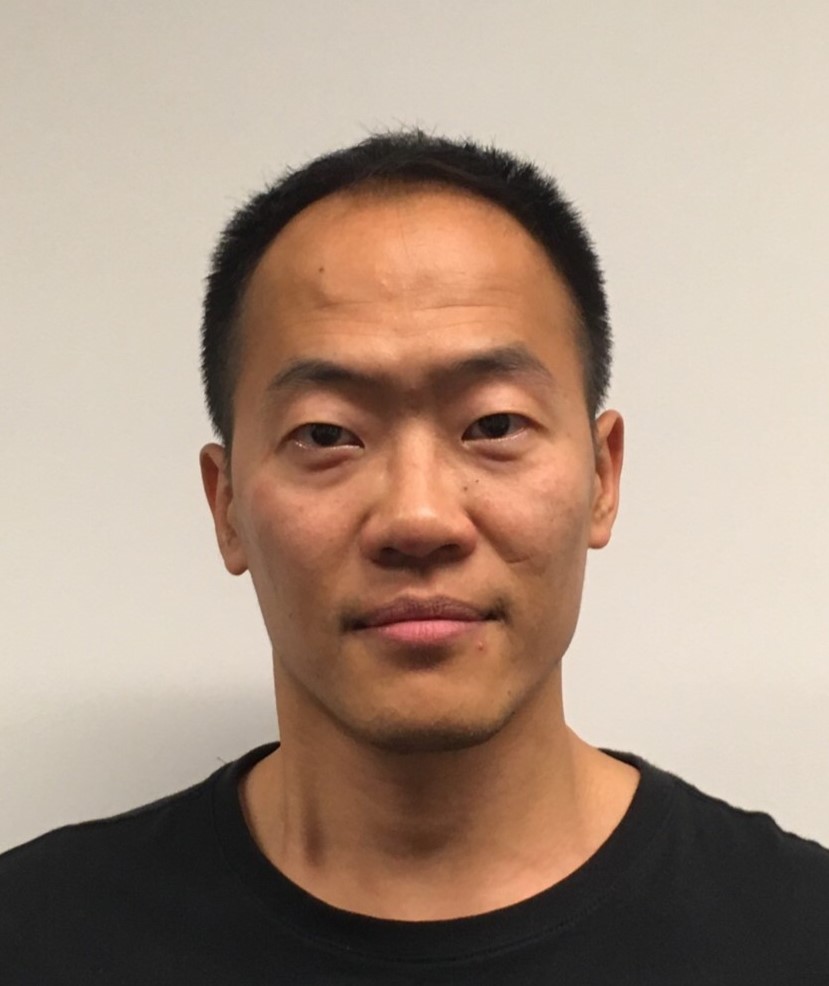}}]{Mingshan Jia} is a Lecturer at the School of Computer Science, University of Technology Sydney. He received a BE degree in information engineering from Xi’an Jiaotong University, China, in 2008, an ME degree in information and telecommunication systems from the University of Technology of Troyes, France, in 2011, and a PhD degree from the University of Technology Sydney, in 2022. Controllability of networks, data mining and machine learning applications in social networks belong to his main interests.
\end{IEEEbiography}
\vskip -2\baselineskip plus -1fil

\begin{IEEEbiography}[{\includegraphics[width=1in,height=1.25in,clip,keepaspectratio]{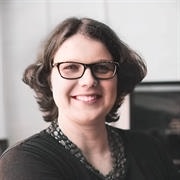}}]{Katarzyna Musial}
received the MSc degree in computer science from the Wroc{\l}aw University of Science and Technology in 2006, the MSc degree in software engineering from the Blekinge Institute of Technology, Sweden, in 2006, and the Ph.D. degree from the Wroc{\l}aw University of Science and Technology, in 2009. In the same year, she was appointed as a Senior Visiting Research Fellow with Bournemouth University, where she has been a Lecturer in informatics since 2010. She joined King’s College London as a Lecturer in computer science in 2011. In 2015, she returned to Bournemouth University where she was an Associate Professor of computing as well as the Head of the SMART Technology Research Group and a member of the Data Science Initiative. In 2017, she moved to Australia and started working as a Professor of Network Science with the Data Science Institute, University of Technology Sydney, where she co-leads the Complex Adaptive Systems Lab. She is particularly interested in social networks, graph controllability and complex systems.
\end{IEEEbiography}
\vskip -2\baselineskip plus -1fil

\begin{IEEEbiography}
[{\includegraphics[width=1in,height=1.25in,clip,keepaspectratio]{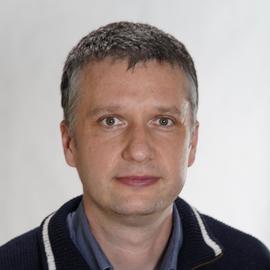}}]{Rados{\l}aw Michalski} is an Associate Professor with the Department of Artificial Intelligence, Wroc{\l}aw University of Science and Technology where he co-leads the Network Science Lab and leads the Blockchain Exploration Research Group (BERG). His research interests include social influence, diffusion processes in complex networks, and machine learning. He has co-authored more than 50 publications in these areas.
\end{IEEEbiography}
\vskip -2\baselineskip plus -1fil

\begin{IEEEbiography}[{\includegraphics[width=1in,height=1.25in,clip,keepaspectratio]{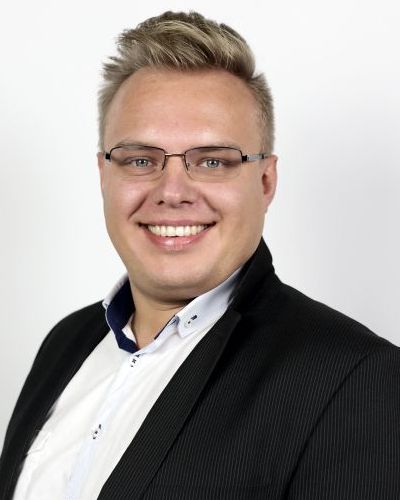}}]{Piotr Br{\'o}dka}
is an Associate Professor at the Department of Artificial Intelligence, Wroc{\l}aw University of Science and Technology. He received an MSc in Computer Science from the Wroc{\l}aw University of Technology in 2008 and a PhD in late 2012. In 2012, he also received an MSc in Computer Science from Blekinge Institute of Technology, Sweden. In 2020, he received a Habilitation (DSc) in Information and Communication Technology. He was a Visiting Scholar at Stanford University in 2013 and a Visiting Professor at the University of Technology Sydney in 2018 and 2019. He has authored over 100 scholarly and research articles on a variety of areas related to complex networks and computational network science, focusing on the extraction and dynamics of communities within social networks, spreading processes in complex networks and the analysis of multilayer networks. 
\end{IEEEbiography}
\vskip -2\baselineskip plus -1fil

\vfill

\end{document}